\documentclass[a4paper,12pt]{article}
\usepackage{amsmath,amsfonts,amssymb}
\usepackage{graphicx}

\def\unit{\relax{\rm 1\kern-.26em I}}

\title{}
\begin{document}

\thispagestyle{empty}

\vspace{-2cm}

\begin{flushright}
{\small CPHT-RR-049.0708 \\
LPT-ORSAY 08-64 \\
SACLAY-T08-110 } \\
\end{flushright}
\vspace{1cm}

\begin{center}
{\bf\LARGE  A light neutralino in hybrid models of supersymmetry breaking}

\vspace{0.8cm}

{\bf Emilian Dudas}$^{a,b}$, {\bf St\'ephane  Lavignac}$^{c}$
{\bf and Jeanne Parmentier}$^{a,c}$
 \vspace{0.8cm}

{\it $^{a}$ Centre de Physique Th\'eorique\! \footnote{Unit\'e mixte
du CNRS (UMR 7644).},
Ecole Polytechnique,\\
F-91128 Palaiseau, France.\\[10pt]
$^{b}$ Laboratoire de Physique Th\'eorique\! \footnote{Unit\'e mixte du CNRS (UMR 8627).},
Universit\'e de Paris-Sud,\\
B\^at. 210, F-91405 Orsay, France.\\ [10pt]
$^{c}$ Institut de Physique Th\'eorique\! \footnote{Laboratoire
de la Direction des Sciences de la Mati\`ere du Commissariat \`a l'Energie
Atomique et Unit\'e de Recherche associ\'ee au CNRS (URA 2306).},
CEA-Saclay,\\
F-91191 Gif-sur-Yvette Cedex, France.

}

\vspace{0.7cm}

{\bf Abstract}
\end{center}
\vspace{.2cm} We show that in gauge mediation models where heavy
messenger masses are provided by the adjoint Higgs field of an underlying
$SU(5)$ theory, a generalized gauge mediation spectrum arises with
the characteristic feature of having a neutralino LSP much lighter
than in the standard gauge or gravity mediation schemes.
This naturally fits in a hybrid scenario where gravity
mediation, while subdominant with respect to gauge mediation,
provides $\mu$ and $B \mu$ parameters of the appropriate size
for electroweak symmetry breaking.

\clearpage

\tableofcontents

\vfill
\eject

\section{Introduction and motivations}

Supersymmetry (SUSY) breaking is the central open question
in supersymmetric extensions of the Standard Model. There are
two major transmission mechanisms, each having its own advantages
and disadvantages:

- gravity mediation~\cite{gravity} easily generates all soft terms
needed at low energy in the Minimal Supersymmetric Standard Model
(MSSM), including the $\mu$ and $B \mu$ terms
of the Higgs sector~\cite{gm}, all being of the order of the
gravitino mass at high energy. A traditional problem
is that the flavor universality needed in order to avoid flavor changing
neutral current (FCNC) transitions is not automatic.
The lightest supersymmetric particle (LSP) in gravity mediation is
generically the lightest neutralino.

- gauge mediation (GMSB)~\cite{gauge1,gauge2,gr} uses Standard
Model gauge loops, and therefore successfully addresses the flavor
problem of supersymmetric models. The soft terms are typically of the order
of a scale determined by the SUSY breaking times a loop factor, which we
call $M_{GM}$ in the following. There is however a serious problem
in generating $\mu$ and $B \mu$ of the right size~\cite{dgp}.
The gravitino, whose mass $m_{3/2}$ is much smaller than $M_{GM}$,
is the LSP. Its lightness is the main signature of gauge mediation.

An obvious way of combining the advantages and possibly reducing
the disadvantages of both mechanisms is to assume
\begin{equation}
  m_{3/2}\ \sim\ (0.01-0.1)\, M_{GM}\ , \qquad M_{GM} \sim 1\, \mbox{TeV} \ .
\label{i1}
\end{equation}
In this case, the FCNC amplitudes induced by the non-universal gravity
contributions to soft scalar masses are suppressed by a factor of order
$m_{3/2}^2/ M_{GM}^2$.
Concerning the $\mu$/$B \mu$ problem,
an option would be to generate $\mu \sim B \sim m_{3/2}$, through
the Giudice-Masiero mechanism~\cite{gm}. However, since
$M_{GM} \gg m_{3/2}$, the squark and gluino masses are much
larger than $m_{3/2}$, and therefore electroweak symmetry breaking 
requires $\mu \gg m_{3/2}$. As we will see explicitly later on,
there is a way of generating $\mu \sim M_{GM}$ in the scenario
considered in this paper, namely through Planck-suppressed
non-renormalizable operators.

Combining the gauge and gravity mediation mechanisms
is an obvious possibility, which has been considered in the past
or more recently from various perspectives~\cite{hybrid}.
It is easy to see that such a hybrid scenario arises for messenger
masses close to the GUT scale. Indeed, consider a set of messenger
fields generically denoted by ($\Phi$, $\tilde \Phi$) coupling
to a set of SUSY breaking fields, generically denoted by $X$:
\begin{equation}
  W_m\ =\ \Phi \left(\lambda_X X + m \right) {\tilde \Phi}\ ,
\label{i2}
\end{equation}
with $\langle X \rangle = X_0 + F_X \theta^2$.
The gauge-mediated contributions to the MSSM soft terms
are proportional to
\begin{equation}
  M_{GM}  \ = \ \frac{g^2}{16 \pi^2}\ \frac{\lambda_X F_X}{M} \ ,
\label{i3}
\end{equation}
where $M = \lambda_X X_0 + m$, and
${g^2 / 16 \pi^2}$ is the loop suppression of gauge mediation.
Since the gravitino mass is given by $m_{3/2} \sim F_X / M_P$
(numerical factors are omitted in this introductory part), the ratio
of the gauge to the gravity contribution reads
\begin{equation}
  \frac{M_{GM}}{m_{3/2}} \ \sim \  \frac{g^2}{16 \pi^2} \lambda_X \frac{M_P}{M} \ ,
\label{i4}
\end{equation}
which shows that gravity mediation is subdominant for
$M \lesssim \frac{g^2}{16 \pi^2}\, \lambda_X M_P \sim \lambda_X M_{GUT}$,
but not completely negligible if $M$ lies within a few orders of magnitude
of $\lambda_X M_{GUT}$.

In the case where messengers come into vector-like pairs of complete
$SU(5)$ multiplets, such as ($5$, $\bar 5$) or ($10$, $\overline{10}$),
and ignoring for simplicity a possible ``flavor'' structure in the messenger
indices, the messenger mass matrix can be written as
\begin{equation}
  M (X) \ = \ \lambda_X X + m \ , \qquad
  m\ =\ m_0\, \unit + \lambda_{\Sigma}\, \langle \Sigma \rangle + \ldots \ ,
\label{i5}
\end{equation}
where $\Sigma$ is the $SU(5)$ adjoint Higgs field. Indeed, any vector-like
pair of complete SU(5) multiplets, besides having an $SU(5)$ symmetric mass
$m_0$, can also couple to $\Sigma$ and get an $SU(5)$-breaking mass term
from its vev. Depending on the messenger
representation, $m$ could also receive contributions from other operators,
denoted by dots in Eq.~(\ref{i5}): operators involving other $SU(5)$
Higgs representations than $\Sigma$, or higher-dimensional operators
such as $\Phi \Sigma^2 \tilde \Phi / M_P$.

From a model-building perspective, the main novelty of the present
paper\footnote{Preliminary results of this paper were reported
at several conferences~\cite{seminars}.} is to consider the case
where the messenger mass matrix is mostly given by the second
term in $m$, i.e. we assume
\begin{equation}
  M(X)\ =\ \lambda_X X + \lambda_\Sigma \langle \Sigma \rangle\ , \quad
  \mbox{with} \quad \lambda_X X_0 \ll \lambda_\Sigma \langle \Sigma \rangle\ .
\label{i5bis}
\end{equation}
As we shall see in Section~\ref{subsec:coupling}, the latter condition
is naturally satisfied when
$X$ is identified with the SUSY breaking field of a hidden sector,
e.g. when $X$ is the meson field of the ISS model~\cite{iss}.
Since\footnote{In the following, we define the $SU(5)$ breaking vev
$v$ by $\langle \Sigma \rangle = v\, \mbox{Diag}\, (2,2,2,-3,-3)$.
By identifying the mass of the superheavy $SU(5)$ gauge bosons
with the scale $M_{GUT}$ at which gauge couplings unify,
we obtain $v = \sqrt{2/25}\, M_{GUT} / g_{GUT} \approx
10^{16}$ GeV.} $\langle \Sigma \rangle = 6 v Y$,  where
$v \approx 10^{16}$ GeV and $Y$ is the hypercharge
generator embedded in $SU(5)$, Eq.~(\ref{i5bis}) implies
\begin{equation}
  M \ = \ 6 \lambda_{\Sigma} v\, Y\ ,
\label{i6}
\end{equation}
up to small corrections of order $\lambda_X X_0$.
Eq.~(\ref{i6}) has a significant impact on the structure of the GMSB-induced
soft terms in the visible (MSSM) sector. Most notably, since the gaugino masses
$M_a$ (where $a$ refers to the SM gauge group factor $G_a = SU(3)_C$,
$SU(2)_L$ or $U(1)_Y$) are proportional to $\mbox{Tr} \left( Q_a^2 / M \right)$,
where the $Q_a$'s stand for the charges of the messenger fields
under $G_a$, it is readily seen that the gauge-mediated contribution
to the bino mass vanishes in the limit $X_0=0$:
\begin{equation}
  \left. M_1 \right|_{GMSB,\, X_0=0}\ \propto\ \mbox{Tr} \left(Y^2 M^{-1} \right)\
  \propto\ \mbox{Tr}\, Y = 0\ .
\label{i7}
\end{equation}
This result holds independently of the $SU(5)$ representation
of the messengers.
A nonzero bino mass is generated from gravity mediation,
from $X_0 \neq 0$ and from possible other terms in $m$,
but it is expected to be much smaller than the other gaugino masses,
which are of order $M_{GM}$. The resulting mass hierarchy,
\begin{equation}
  M_1\ \ll\ M_2 \sim M_3 \sim \mu\ ,
\end{equation}
leads to a light mostly-bino neutralino, which is therefore the LSP
(unless $M_1 \gtrsim 2 m_{3/2}$ at the messenger scale, in which
case the LSP is the gravitino). In addition to being theoretically
well motivated, this scheme provides
a natural realization of the light neutralino scenarios occasionally
considered in the litterature~\cite{Dreiner,HP02,BFS02,BBCPR03,BFPS08},
and invoked more recently~\cite{dama} in connection with the new
DAMA/LIBRA data~\cite{damaresult}.

The plan of the paper is the following. In Section~\ref{sec1}, we
present the MSSM soft terms induced by the messenger mass
matrix~(\ref{i5bis}), which breaks the $SU(5)$ symmetry in a well-defined
manner. In Section~\ref{sec2}, we couple the messenger sector
to an explicit (ISS) supersymmetry breaking sector.  We study the stability
of the phenomenologically viable vacuum after including quantum
corrections, and discuss the generation of the $\mu$ and $B \mu$
terms by Planck-suppressed operators. In Section~\ref{sec:pheno},
we discuss the low-energy phenomenology of the scenario, paying
particular attention to the dark matter constraint. Finally,
we present our conclusions in Section~\ref{sec:conclusions}.
The appendices contain technical details about the computation
of the MSSM soft terms and the quantum corrections to the scalar
potential.

\section{ Gauge mediation with GUT-induced messenger mass splitting}
\label{sec1}

The main difference between minimal gauge mediation and
the scenario considered in this paper\footnote{For recent analyses
of general messenger masses, see e.g. Refs.~\cite{cfs}.} lies
in the messenger mass matrix~(\ref{i5bis}). The messenger mass splitting
depends on the $SU(5)$ representation of the messenger fields.
Denoting by $(\phi_i, \tilde \phi_i)$ the component messenger fields
belonging to definite SM gauge representations and by $Y_i$ their
hypercharge, one has
\begin{equation}
  \mbox{Tr}\, (\Phi \langle \Sigma \rangle {\tilde \Phi})\ =\
  6 v\, \sum_i\, Y_i\, \phi_i \tilde \phi_i \ ,
\label{mssm1}
\end{equation}
yielding a mass $M_i = 6 \lambda_\Sigma v Y_i$ for $(\phi_i, \tilde \phi_i)$
(again $X_0 = 0$ is assumed). In the cases of $(\bar 5, 5)$ and $(10, \overline{10})$
messengers, the component fields and their masses are, respectively,
\begin{eqnarray}
&  \Phi (\bar 5)\, =\, \left\{ \phi_{\bar 3,1,1/3}\, ,\, \phi_{1,2,-1/2} \right\} , \qquad
  \qquad M\, =\, \left\{ 2 \lambda_\Sigma v\, ,\, - 3 \lambda_\Sigma v \right\} ,  \\
& \Phi (10)\, =\, \left\{ \phi_{3,2,1/6}\, ,\, \phi_{{\bar 3},1,-2/3}\, ,\,
  \phi_{1,1,1} \right\} ,  \quad M\, =\, \left\{ \lambda_\Sigma v\, ,\,
  - 4 \lambda_\Sigma v\, ,\, 6 \lambda_\Sigma v \right\} ,
\label{mssm2}
\end{eqnarray}
where the subscripts denote the $SU(3)_C \times SU(2)_L \times U(1)_Y$
quantum numbers, and the components ${\tilde \phi_i}$ of $\tilde \Phi$ are
in the complex conjugate representations.

The one-loop GMSB-induced gaugino masses are given by
(see Appendix~\ref{app1})
\begin{equation}
  M_{a} (\mu) \ = \ \frac{\alpha_a(\mu)}{4\pi}\ \sum_i\, 2 T_a (R_i)\,
  \frac{\partial \ln (\det M_i )}{\partial \ln X} \left. \frac{F_X}{X}\, \right|_{X=X_0}\ ,
\label{mssm4}
\end{equation}
where
the sum runs over the component messenger fields $(\phi_i, \tilde \phi_i)$,
and $T_a(R_i)$ is the Dynkin index of the representation $R_i$ of $\phi_i$.
As noted in the introduction, with the messenger mass matrix~(\ref{i5bis}),
the gauge-mediated contribution to the bino mass vanishes
irrespective of the  $SU(5)$ representation of the messengers,
up to a correction proportional to  $\lambda_X X_0$ which will turn out
to be negligible (see Section~\ref{subsec:coupling}). Then $M_1$
is mainly of gravitational origin:
\begin{equation}
  M_1\ \sim\ m_{3/2}\ .
\label{mssm5}
\end{equation}

As the messenger masses are not $SU(5)$ symmetric, the running
between the different messenger scales should be taken into account
in the computation of the soft scalar masses. The corresponding formulae
are given in Appendix~\ref{app1}. For simplicity, we write below the
simpler expressions
obtained when the effect of this running is neglected. The two-loop
MSSM soft scalar mass parameter $m^2_\chi$,
induced by $N_1$ messengers of mass $M_1$ and $N_2$
messengers of mass $M_2$ and evaluated at the messenger scale, reads
\begin{equation}
  m^2_\chi\ =\ 2 \sum_a C^a_\chi \left( \frac{\alpha_a}{4\pi} \right)^2\!
    \biggl\{ 2 N_2 T_a(R_2) \left| \frac{\partial \ln M_2}{\partial \ln X} \right|^2\!
    + 2 N_1 T_a(R_1) \left| \frac{\partial \ln M_1}{\partial \ln X} \right|^2 \biggr\}
    \left| \frac{F_X}{X} \right|^2  . 
\label{mssm6}
\end{equation}
In Eq.~(\ref{mssm6}), $C^a_\chi$ are the second Casimir coefficients
for the superfield $\chi$, normalized to $C (N) = (N^2-1)/2N$ for the
fundamental representation of $SU(N)$ and to $C^1_\chi = 3 Y^2_\chi / 5$
for $U(1)$, and $T_a(R_i)$ are the Dynkin indices for the messenger fields.

While the vanishing of the GMSB contribution to the bino mass is a
simple consequence of the underlying hypercharge embedding in a simple
gauge group and of the structure of the mass matrix~(\ref{i5bis}) (i.e.
it is independent of the representation of the messengers), the
ratios of the other superpartner masses, including the ratio of the gluino
to wino masses $M_3/M_2$, do depend on the representation
of the messengers. This is to be compared with minimal gauge
mediation~\cite{gr}, in which the ratios of gaugino masses (namely,
$M_1 : M_2 : M_3 = \alpha_1 : \alpha_2 : \alpha_3$) as well as the ratios
of the different scalar masses is independent of the
representation of the messengers~\cite{gaugepheno}. Leaving a
more extensive discussion of the mass spectrum to Section~\ref{sec:pheno},
we exemplify this point below with the computation of the gaugino and
scalar masses in the cases of $(5, \bar 5)$ and $(10, \overline{10})$
messengers:

{\bf i) $(5,{\bar 5})$ messenger pairs:} in this case the gluino
and $SU(2)_L$ gaugino masses are given by
\begin{equation}
M_3 \ = \ \frac{1}{2}\, N_m\, \frac{\alpha_3}{4 \pi}\, \frac{\lambda_X F_X}
  {\lambda_{\Sigma} v}\ , \qquad M_2 \ = \ - \frac{1}{3}\, N_m\,
  \frac{\alpha_2}{4 \pi}\, \frac{\lambda_X F_X}{\lambda_{\Sigma} v} \ ,
\label{eq:gauginos_5}
\end{equation}
where $N_m$ is the number of messenger pairs,
leading to the ratio $|M _3/M_2| = 3 \alpha_3 /2 \alpha_2$
($\approx 4$ at $\mu = 1$ TeV). The complete expressions for
the scalar masses can be found in Appendix~\ref{app1}.
For illustration, we give below the sfermion
soft masses at a messenger scale of $10^{13}$ GeV,
neglecting the running between the different messenger
mass scales as in Eq.~(\ref{mssm6}):
\begin{equation}
  m^2_Q\, :\, m^2_{U^c}\, :\, m^2_{D^c}\, :\, m^2_L\, :\, m^2_{E^c}\ \approx\
  0.79\, :\, 0.70\, :\, 0.68\, :\, 0.14\, :\, 0.08\ ,
\label{eq:scalars_5}
\end{equation}
in units of $N_m M^2_{GM}$, with
$M_{GM} \equiv (\alpha_3 / 4 \pi) (\lambda_X F_X / \lambda_\Sigma v)$.
In Eq.~(\ref{eq:scalars_5}) as well as in Eq.~(\ref{eq:scalars_10})
below, we used $(\alpha_1 / \alpha_3) (10^{13}\, \mbox{GeV}) = 0.65$
and $(\alpha_2 / \alpha_3) (10^{13}\, \mbox{GeV}) = 0.85$.

{\bf ii) $(10, \overline{10})$ messenger pairs:} in this case the gluino and $SU(2)_L$
gaugino masses are given by
\begin{equation}
M_3 \ = \ \frac{7}{4}\, N_m\, \frac{\alpha_3}{4 \pi}\, \frac{\lambda_X F_X}
  {\lambda_{\Sigma} v}\ , \qquad M_2 \ = \ 3\, N_m\,
  \frac{\alpha_2}{4 \pi}\, \frac{\lambda_X F_X}{\lambda_{\Sigma} v} \ ,
\label{eq:gauginos_10}
\end{equation}
leading to the ratio $M_3/M_2 = 7 \alpha_3 / 12 \alpha_2$
($\approx 1.5$ at $\mu = 1$ TeV). In this case too, we give the sfermion
soft masses at a messenger scale of $10^{13}$ GeV for illustration:
\begin{equation}
  m^2_Q\, :\, m^2_{U^c}\, :\, m^2_{D^c}\, :\, m^2_L\, :\, m^2_{E^c}\ \approx\
  8.8\, :\, 5.6\, :\, 5.5\, :\, 3.3\, :\, 0.17\ ,
\label{eq:scalars_10}
\end{equation}
again in units of $N_m M^2_{GM}$. For the Higgs soft masses,
one has $m^2_{H_u} = m^2_{H_d} = m^2_L$ irrespective
of the messenger representation.

In contrast to minimal gauge mediation with $SU(5)$ symmetric
messenger masses, in which the ratios of gaugino masses
are independent of the messenger representation, in our scenario
the gaugino mass ratios and more generally the detailed MSSM
mass spectrum are representation dependent. There is however one
clear-cut prediction, which distinguishes it from both
minimal gauge mediation and minimal gravity mediation,
namely the vanishing of the one-loop GMSB contribution to the bino mass.
Notice also the lightness of the scalar partners of the right-handed leptons
for ($10$, $\overline{10}$) messengers, which arises from
the correlation between the hypercharge and the mass
of the different component messenger fields
(the lightest components have the smallest hypercharge).
Finally, we would like to point out that, due to the fact that messengers
carrying different SM gauge quantum numbers have different masses,
gauge coupling unification is slightly modified compared to the MSSM.
Since the messengers are heavy and their mass splitting is not very
important, however, this effect is numerically small.

In the above discussion, the higher-dimensional operator
$\lambda'_\Sigma\, \Phi \Sigma^2 \tilde \Phi / M_P$ was assumed
to be absent. Before closing this section, let us briefly discuss
what relaxing this assumption would imply. If $\lambda'_\Sigma \neq 0$,
the messenger mass matrix~(\ref{i6}) receives an additional
contribution, which affects the gauge-mediated MSSM soft terms.
In particular, $M_1$ no longer vanishes:
\begin{equation}
  \left. M_1 \right|_{GMSB}\ =\ - \frac{6}{5}\, d\, \mbox{Tr}\, Y^2\,
  \frac{\lambda'_\Sigma v}{\lambda_\Sigma M_P}\,
  \frac{\alpha_1}{\alpha_3}\ N_m M_{GM}\ ,
\label{eq:DeltaM1}
\end{equation}
where $d$ is the dimension of the messenger representation,
and the trace is taken over the representation. Eq.~(\ref{eq:DeltaM1})
was derived under the assumption that the $\lambda'_\Sigma$-induced
corrections to the messenger masses are small, so that to a good
approximation, the scalar and electroweak gaugino masses are still
given by Eqs.~(\ref{eq:gauginos_5}) to~(\ref{eq:scalars_10}).
It is easy to show that this implies
\begin{equation}
  \left. M_1 \right|_{GMSB}\ \ll\ 0.2\, N_m M_{GM}\ ,
\end{equation}
for both $(5, \bar 5)$ and $(10, \overline{10})$ messengers.
In the rest of the paper, we shall therefore neglect the contribution
of $\lambda'_\Sigma \neq 0$ and assume that $M_1$ is generated
by gravity mediation.

\section{A complete model }
\label{sec2}

The computation of the MSSM soft terms performed in the previous
section is to a large extent insensitive to the details of the
sector that breaks supersymmetry. The generation of the $\mu$ and
$B \mu$ terms, on the other hand, depends
on its details. The goal of the present section is to consider an
explicit SUSY breaking sector, to couple it to the messenger sector,
and to check that the following constraints are satisfied:
(i) nonperturbative instabilities towards possible color-breaking
vacua are sufficiently suppressed; (ii) $\mu$ and $B \mu$ parameters
of the appropriate size can be generated.

The model can be described by a superpotential of the form:
\begin{equation}
W \ = \ W_{MSSM} \ + \ W_{SB} (X, \ldots) \ + \ W_m (\Phi, {\tilde \Phi}, X, \Sigma) \
  + \ W_{GUT} (\Sigma) \ ,
\label{complete1}
\end{equation}
where $W_{SB} (X, \ldots)$ describes the SUSY breaking sector,
$W_m (\Phi, {\tilde  \Phi}, X, \Sigma)$ the couplings of the messengers
fields ($\Phi, \tilde \Phi$) to the SUSY breaking fields $X$
and to the $SU(5)$ adjoint Higgs field $\Sigma$,
and $W_{GUT} (\Sigma)$
describes the breaking of the unified gauge symmetry,
$SU(5) \rightarrow SU(3)_C \times SU(2)_L \times U(1)_Y$.
In this paper, we consider the case where
$W_m (\Phi, {\tilde  \Phi}, X, \Sigma)
= \Phi \left( \lambda_X X + \lambda_\Sigma \Sigma \right) \tilde \Phi$.
The details of the GUT sector are irrelevant for our purposes
and will not be further discussed in the following. The implicit assumption
here is that the SUSY breaking sector and the GUT sector only couple
via gravity and via the messenger fields. It is therefore reasonable
to expect that they do not influence significantly their respective dynamics.

\subsection{The SUSY breaking sector}

A generic dynamical supersymmetry breaking sector~\cite{dsb} coupled to
the messenger sector is enough for our purposes. For concreteness
and simplicity, we consider here the ISS model~\cite{iss}, namely
${\cal N}=1$ SUSY QCD with $N_f$ quark flavors and gauge
group $SU(N_c)$ in the regime $N_c < N_f < \frac{3}{2}\, N_c$.
In the IR, the theory is strongly coupled, giving rise to a low-energy
physics that is better described by a dual ``magnetic'' theory
with gauge group $SU(N_f-N_c)$, $N_f$ flavors of quarks $q^i_a$
and antiquarks $\tilde q_i^a$, and meson (gauge singlet) fields
$X_i^j$ ($i,j = 1 \ldots N_f$, $a = 1\ldots N$, with $N \equiv N_f - N_c$).
The magnetic theory is IR free and can be analyzed perturbatively.

The superpotential of the magnetic theory,
\begin{equation}
  W_{ISS} \ = \ h\, q^i_a X_i^j \tilde{q}_j^a \ - \ h f^2\, \mbox{Tr} X\ ,
\label{iss1}
\end{equation}
leads to supersymmetry breaking a la O'Raifeartaigh,
since the auxiliary fields
$(-F^\star_X)^i_j = h q^i_a \tilde{q}_j^a - h f^2 \delta^i_j $ cannot all
be set to zero. Indeed, the matrix $q^i_a \tilde{q}_j^a$ is at most
of rank $N$, whereas the second term $h f^2 \delta^i_j$ has
rank $N_f > N$.
The supersymmetry-breaking ISS vacuum is defined by
$\langle q^i_a \rangle = \langle \tilde q_i^a \rangle = f \delta_i^a$,
$\langle X \rangle = 0$. At tree level, there are flat directions along
which the components $i, j = (N+1) \ldots N_f$ of $X_i^j$ are non-vanishing;
quantum corrections lift
them and impose $ \langle X \rangle = 0$~\cite{iss}. This means
that the R-symmetry under which X is charged is not spontaneously
broken, which in turn implies that no gaugino masses are generated
in the minimal ISS model.
Another important feature of the ISS vacuum is that it is metastable.
Indeed, according to the Witten
index, the theory possesses $N_c$ supersymmetric vacua. These
vacua are obtained in the magnetic description by going along the branch
with nonzero meson vev's, $\langle X \rangle \neq 0$, where magnetic
quarks become massive and decouple, so that the low-energy theory
becomes strongly coupled again. In order to ensure that the lifetime
of the ISS vacuum is larger than the age of the universe, one
requires $f \ll \Lambda_m$, where $\Lambda_m$ is the scale
above which the magnetic theory is strongly coupled.

\subsection{Coupling the SUSY breaking sector to messengers}
\label{subsec:coupling}

Let us now couple the SUSY breaking sector to the messenger sector
by switching on the superpotential term $\lambda_X \Phi X \tilde \Phi$,
and address the following two questions:
\begin{itemize}
\item How is the vacuum structure of the model affected, in particular
is the ISS vacuum still metastable and long lived?
\item Is it possible to generate $\mu$ and $B \mu$ of the appropriate size?
\end{itemize}
The first question has been investigated in several works~\cite{gaugeiss}
in the case of $SU(5)$ symmetric messenger masses. We reanalyze it
in our scenario and come to a similar conclusion: the messenger fields
induce a lower minimum which breaks the SM gauge symmetries, a rather
common feature of gauge mediation models. To our knowledge, the solution
we propose for the second issue has not been discussed in the
literature\footnote{For recent approaches to the $\mu$/$B \mu$ problem
of gauge mediation, see Refs.~\cite{mugauge}.}. We now proceed
to address the above two questions in detail.

\subsubsection{Stability of the phenomenologically viable vacuum}

It is well known that coupling a SUSY breaking sector to a messenger
sector generally introduces lower minima in which the messenger fields
have nonzero vev's. Since the messengers carry SM gauge quantum
numbers, these vacua are phenomenologically unacceptable.
Such minima also appear in our scenario. Summarizing the analysis
done in Appendix B, we indeed find two types of
local supersymmetry-breaking minima at tree level:
\begin{itemize}
\item the ISS vacuum with
  no messenger vev's and energy
  $V(\phi \tilde{\phi} = 0 )= (N_f-N) h^2 f^4$;
\item lower minima with messenger vev's $\phi \tilde{\phi}\ =\
  \frac{- \displaystyle \sum_{i=N+1}^{N_f} \lambda_{X,i}^i h f^2}
  {\displaystyle \sum_{(i,j) \notin \{i=j=1 \ldots N \}} \left| \lambda_{X,j}^i \right|^2}$ \
  and energy \\
$V (\phi \tilde{\phi} \neq 0)\ =\ h^2 f^4 \left( N_f - N\
  -\ \left| \displaystyle\sum_{i=N+1}^{N_f} \lambda_{X,i}^i\right|^2 \left/ \! \!
  \displaystyle \sum_{(i,j) \notin \{i=j=1 \ldots N \}} \left| \lambda_{X,j}^i \right|^2 \right. \right) .$
\end{itemize}
Transitions from the phenomenologically viable ISS minimum to the
second class of minima, in which the SM gauge symmetry is broken
by the messenger vev's, must be suppressed.
An estimate of the lifetime of the ISS vacuum in the triangular
approximation gives  $\tau \sim \exp(\frac{(\Delta \phi)^4}{\Delta V}) $, with
\begin{equation}
  \frac{\Delta V}{(\Delta \phi)^4}\ \ =\ \sum_{(i,j) \notin \{i=j=1 \ldots N \}} |\lambda_{X,j}^i|^2\ \
  \equiv\ \ \overline{\lambda^2}\ .
\end{equation}
The lifetime of the phenomenologically viable vacuum is therefore
proportional to $e^{1/\overline{\lambda^2}}$.
To ensure that it is larger than the age of the universe, it is enough
to have $\overline{\lambda^2} \lesssim 10^{-3}$.

We conclude that, as anticipated, the superpotential coupling
$\lambda_X \Phi X \tilde \Phi$ induces new minima with a lower energy
than the ISS vacuum, in which the messenger fields acquire vev's
that break the SM gauge symmetry. 
In order to ensure that the ISS vacuum is sufficiently long lived, the coupling
between the ISS sector and the messenger sector, $\lambda_X$,
has to be small. We believe that this result is quite generic.

Let us now discuss the stability of the phenomenologically viable vacuum
under quantum corrections. As shown in Ref.~\cite{iss},
the ISS model possesses tree-level flat directions
that are lifted by quantum corrections. The novelty of our analysis
with respect to Ref.~\cite{iss} is that we include messenger loops
in the computation of the one-loop effective potential, and we find that
these corrections result in a nonzero vev for $X$. The detailed
analysis is given in Appendix~\ref{app2}; here we just notice that since
the messenger fields do not respect the R-symmetry of the ISS sector,
it is not surprising that coupling the two sectors induces a nonzero vev
for $X$ (which otherwise would be forbidden by the R-symmetry).
Indeed, the one-loop effective potential for the meson fields reads,
keeping only the leading terms relevant for the minimization procedure
(see Appendix~\ref{app2} for details):
\begin{eqnarray}
&& V_{\rm 1-loop} (X_0, Y_0)\ =\ 2N h^2 f^2 |Y_0|^2\,
  +\, \frac{1}{64 \pi^2}\ \Big\{ 8 h^4 f^2 (\ln 4 -1) N (N_f-N) |X_0|^2 \nonumber \\
&& \qquad \qquad +\ \frac{10 N_m h^2 f^4 | \mbox{Tr}' \lambda |^2}{3 \lambda_{\Sigma} v}\
  \Big[\, (\mbox{Tr}' \lambda)\, X_0\, +\, (\mbox{Tr}'' \lambda)\, Y_0\, +\, {\rm h.c.}\, \Big]
  \Big\}\ ,
\label{stability1}
\end{eqnarray}
where we have set $\tilde X = X_0\, \unit_{N_f-N}$, $\tilde Y = Y_0\, \unit_{N}$
and defined $\mbox{Tr}' \lambda \equiv \sum_{i=N+1}^{N_f} \lambda_{X,i}^i$,
$\mbox{Tr}'' \lambda \equiv \sum_{i=1}^{N} \lambda_{X,i}^i$.
In Eq.~(\ref{stability1}), the first line contains the tree-level potential
for $X$ and the one-loop corrections computed in Ref.~\cite{iss},
whereas the linear terms in the second line are generated
by messenger loops. The latter induce vev's for the meson fields:
\begin{equation}
\langle X_0 \rangle\ \simeq\  -\ \frac{5 N_m\, |\mbox{Tr}' \lambda|^2\,
  (\mbox{Tr}' \lambda)^\star}{12 (\ln 4-1) h^2 N (N_f-N)}\
  \frac{f^2}{\lambda_{\Sigma} v}\ ,
\label{stability01}
\end{equation}
\begin{equation}
\langle Y_0 \rangle\ \simeq\  -\ \frac{5 N_m\, |\mbox{Tr}' \lambda|^2\,
  (\mbox{Tr}'' \lambda)^\star}{192 \pi^2 N}\
  \frac{f^2}{\lambda_{\Sigma} v}\ .
\label{stability02}
\end{equation}
Notice that, due to $\langle Y_0 \rangle \neq 0$, magnetic quarks
(and antiquarks) do contribute to supersymmetry breaking:
$F_q \sim {\tilde q} X \not=0$ ($F_{\tilde q} \sim q X \not=0$),
while $F_q = F_{\tilde q} = 0$ in the ISS model as a consequence
of the R-symmetry. Here instead,  the R-symmetry is broken
by the coupling of the ISS sector to the messengers fields,
and the F-terms of the magnetic (anti-)quarks no longer
vanish.
We have checked that, in the messenger direction,  $\phi =\tilde{\phi}=0$
is still a local minimum. We have also checked that the nonzero
vev's~(\ref{stability01}) and (\ref{stability02}) resulting from quantum
corrections do not affect the discussion about the lifetime
of the phenomenologically viable vacuum. Notice that these vev's
also appear in the standard case where messenger masses
are $SU(5)$ symmetric.

\subsubsection{Generation of the $\mu$ and $B \mu$ terms}

As stressed in the introduction, due to the hierarchy of scales
$m_{3/2} \ll M_{GM}$, the Giudice-Masiero mechanism fails
to generate a $\mu$ term of the appropriate magnitude
for radiative electroweak symmetry breaking. Fortunately,
there are other sources for $\mu$ and $B \mu$ in our scenario.

A crucial (but standard)
hypothesis is the absence of a direct coupling between the hidden
SUSY breaking sector and the observable sector (i.e. the MSSM).
In particular, the coupling $X H_u H_d$ should be absent from the
superpotential. The fields of the ISS sector therefore couple to the
MSSM fields only via non-renormalizable interactions
and via the messengers. 
It is easy to check that non-renormalizable interactions involving
the ISS and MSSM fields have a significant effect only on the
$\mu$ and $B \mu$ terms, whereas they induce negligible corrections
to the MSSM soft terms and Yukawa couplings.
The most natural operators mixing the two sectors, which are local
both in the electric and in the magnetic phases of the ISS model,
are the ones built from the mesons $X$. It turns out, however,
that such operators do not generate $\mu$ and $B \mu$ parameters
of the appropriate magnitude.

Fortunately, a more interesting possibility arises in our scenario,
thanks to the loop-induced vev of the meson fields discussed
in the previous subsection. Indeed, the Planck-suppressed operator
\begin{equation}
  \lambda_1\, \frac{q \tilde{q}}{M_P}\, H_u H_d\ ,
\label{mu1}
\end{equation}
in spite of being of gravitational origin, yields a $\mu$ term
that can be parametrically larger than $m_{3/2}$.
This allows us to assume
$m_{3/2} \ll M_{GM}$, as needed to suppress the most dangerous
FCNC transitions, consistently with electroweak symmetry breaking
(which typically requires a $\mu$ term of the order of the squark
and gluino masses).
More precisely, the operator~(\ref{mu1}) generates
\begin{equation}
  \mu \ = \  \frac{\lambda_1}{h}\, \frac{N}{\sqrt{N_c}}\, \sqrt{3}\, m_{3/2}\ ,
\label{mu2}
\end{equation}
\begin{equation}
  B \ = \
  - 2 h \langle Y^\star_0 \rangle\
  =\ -\, \frac{5 N_m\, |\mbox{Tr}' \lambda|^2\, (\mbox{Tr}'' \lambda)}{96 \pi^2 N \sqrt{N_c}}\,
  \frac{M_P}{\lambda_{\Sigma} v}\ \sqrt{3}\, m_{3/2}\ ,
\label{mu3}
\end{equation}
where we used
$m_{3/2} = \sqrt{\sum_{i=N+1}^{N_f} |F_{X,i}^i|^2}\, / \sqrt{3} M_P
= \sqrt{N_c}\, h f^2 / \sqrt{3} M_P$.
Using Eqs.~(\ref{stability01}) and~(\ref{stability02}), it is easy to convince
oneself that one can obtain
$\mu \sim 1$~TeV for e.g.
$m_{3/2} \sim (10-100)$ GeV, by taking a small enough ISS coupling $h$.
As a numerical example, one can consider for instance
$m_{3/2} = 50$ GeV, $N_c=5$, $N_f=7$ and $\lambda_1 / h = 10$,
in which case $\mu = 775$ GeV. As for the $B$ parameter, it turns out
to be somewhat smaller than $m_{3/2}$.
For instance, taking as above $N_c=5$, $N_f=7$ and assuming further $N_m = 1$,
$|\mbox{Tr}' \lambda|^2 = 10^{-3}$ and $\lambda_\Sigma v = 10^{13}$ GeV,
one obtains $B = - 0.49\, (\mbox{Tr}'' \lambda)\, m_{3/2}$.
This will in general
be too small for a proper electroweak symmetry breaking, 
even if $\mbox{Tr}'' \lambda \sim 1$ is possible in principle
(contrary to $\mbox{Tr}' \lambda$, $\mbox{Tr}'' \lambda$ is not
constrained by the lifetime of the ISS vacuum).
However, $B \mu$ also
receives a contribution from the non-renormalizable operator
\begin{equation}
  \lambda_2\, \frac{X X}{M_P}\, H_u H_d\ ,
\label{mu4}
\end{equation}
which gives a negligible contribution to $\mu$, but yields
$B \mu = - \lambda_2 \sqrt{3 N_c}\,  \langle X_0 \rangle\, m_{3/2}$.
Using Eq.~(\ref{mu2}), one then obtains
\begin{equation}
  B \ = \
  -\, \lambda_2\, \frac{h}{\lambda_1}\, \frac{N_c}{N}\, \langle X_0 \rangle\
  =\ -\, \frac{\lambda_2}{\lambda_1}\, \frac{5 N_m\, |\mbox{Tr}' \lambda|^2\,
  (\mbox{Tr}' \lambda)^\star}{12 (\ln 4-1) h^2 N^2 \sqrt{N_c}}\,
  \frac{M_P}{\lambda_{\Sigma} v}\ \sqrt{3}\, m_{3/2}\ ,
\label{mu5}
\end{equation}
which is enhanced with respect to Eq.~(\ref{mu3}) by the absence
of the loop factor and by the presence of $h^2$ in the denominator.
It is then easy to obtain the desired value of the $B$ parameter.
As an illustration, choosing the same parameter values as in the
above numerical examples and taking  $h = 0.1$,
one obtains $B / \lambda_2 = 7.9$ TeV, while choosing
$\mbox{Tr}' \lambda = 10^{-2}$ (instead of $10^{-3/2}$)
gives $B / \lambda_2 = 250$ GeV.

We conclude that Planck-suppressed operators can generate
$\mu$ and $B \mu$ parameters of the appropriate size in our scenario,
thanks to the vev's of the meson fields induced by messenger loops,
which are crucial for the generation of $B \mu$.
As mentioned in the previous subsection, these vev's appear
independently of whether the messenger masses are split or not.
Therefore, the $\mu$ and $B \mu$ terms can be generated in the same
way in more standard gauge mediation models with $SU(5)$ symmetric
messenger masses.

Notice that there is a price to pay for the above solution
to the $\mu$/$B\mu$ problem: the interaction term~(\ref{mu1}),
which is local in the magnetic ISS description, becomes non-local
in the electric description, analogously to the $q X {\tilde q}$ coupling
of the magnetic Seiberg duals~\cite{seibergduality}.

\section{Low-energy phenomenology}
\label{sec:pheno}

The phenomenology of minimal gauge mediation has been
investigated in detail in the past (see e.g. Ref.~\cite{gaugepheno}).
The main distinctive feature of our scenario with respect to standard
gauge mediation is the presence of a light neutralino, with a mass
of a few tens of GeV
in the picture where $M_1 \sim m_{3/2} \sim (10-100)$ GeV.
As is well known, such a light neutralino
is not ruled out by LEP data: the usually quoted lower bound
$M_{\tilde \chi^0_1} \gtrsim 50$ GeV
assumes high-scale gaugino mass unification,
and can easily be evaded once this assumption is relaxed\footnote{More
precisely, for a mostly-bino neutralino (as in our scenario,
where $M_1 \ll M_2, |\mu|$), there is no mass bound from LEP if either
$M_{\tilde \chi^0_1} + M_{\tilde \chi^0_2} > 200$ GeV or selectrons are very
heavy~\cite{Drees_book}. The former constraint is satisfied by all superpartner
mass spectra considered in this section. Furthermore, a mostly-bino
neutralino has a suppressed coupling to the $Z$ boson and thus only
gives a small contribution to its invisible decay width.}.
The other features of the
superpartner spectrum depend on the messenger representation.
Particularly striking is the lightness of the $\tilde l_R$ with respect
to other sfermions (including the $\tilde l_L$) in the case of
($10$, $\overline{10}$) messengers.
The values of the soft terms at the reference messenger
scale\footnote{As explained in Appendix~\ref{app2_lifetime}, the
requirement that our metastable vacuum is sufficiently long lived
constrains the messenger scale $M_{mess} \equiv \lambda_\Sigma v$
to lie below $10^{14}$ GeV or so. Demanding $M_{GM} / m_{3/2} \sim 10$
further pushes it down to $10^{13}$ GeV.}
$M_{mess} = 10^{13}$ GeV are given
by Eqs.~(\ref{eq:gauginos_5}) to~(\ref{eq:scalars_10}). One can
derive approximate formulae for the gaugino and the first two generation
sfermion masses at low energy by neglecting the Yukawa contributions
in the one-loop renormalization group equations, as expressed by Eq.~(\ref{a6}).
At the scale $\mu = 1$ TeV, one thus obtains
\begin{equation}
  M_2\, \simeq\, 0.25 N_m M_{GM}\, ,  \quad
  M_3\, \simeq\, N_m M_{GM}\, ,
\label{eq:gauginos_5_LE}
\end{equation}
\vskip -1.2cm
\begin{eqnarray}
&& m^2_{Q_{1,2}}\, \simeq\, (0.79 + 0.69 N_m) N_m M^2_{GM}\, , \quad
  m^2_{U^c_{1,2}}\, \simeq\, (0.70 + 0.66 N_m) N_m M^2_{GM}\, ,  \\
&& m^2_{D^c_{1,2}}\, \simeq\, (0.68 + 0.66 N_m) N_m M^2_{GM}\, ,  \quad
  m^2_{L_{1,2}}\, \simeq\, (0.14 + 0.03 N_m) N_m M^2_{GM}\, ,  \hskip 1cm \\
&& m^2_{E^c_{1,2}}\, \simeq\, 0.08 N_m M^2_{GM} + 0.12 M^2_1\, ,
\label{eq:sfermions_5_LE}
\end{eqnarray}
for $(5, \bar 5)$ messengers, and
\begin{equation}
  M_2\, \simeq\, 2.2 N_m M_{GM}\, ,  \quad
  M_3\, \simeq\, 3.5 N_m M_{GM}\, ,
\label{eq:gauginos_10_LE}
\end{equation}
\vskip -1.2cm
\begin{eqnarray}
&& m^2_{Q_{1,2}}\, \simeq\, (8.8 + 10.4 N_m) N_m M^2_{GM}\, , \quad
  m^2_{U^c_{1,2}}\, \simeq\, (5.6 + 8.1 N_m) N_m M^2_{GM}\, ,  \\
&& m^2_{D^c_{1,2}}\, \simeq\, (5.5 + 8.1 N_m) N_m M^2_{GM}\, ,  \quad
  m^2_{L_{1,2}}\, \simeq\, (3.3 + 2.3 N_m) N_m M^2_{GM}\, ,  \\
&& m^2_{E^c_{1,2}}\, \simeq\, 0.17 N_m M^2_{GM} + 0.12 M^2_1\, ,
\label{eq:sfermions_10_LE}
\end{eqnarray}
for ($10$,$\overline{10}$) messengers, where
$M_{GM} = (\alpha_3 (M_{mess}) / 4 \pi) (\lambda_X F_X / \lambda_\Sigma v)$.
Furthermore, one has in both cases:
\begin{equation}
  M_{\tilde \chi^0_1}\ \approx\ 0.5 M_1\ .
\label{eq:M_chi0_1}
\end{equation}
In Eqs.~(\ref{eq:gauginos_5_LE}) to~(\ref{eq:sfermions_10_LE}),
the unknown gravitational contribution to the soft terms is not taken
into account, apart from $M_1$ which is taken as an input (we neglected
subdominant terms proportional to $M^2_1$ in all sfermion masses
but $m^2_{E^c_{1,2}}$). 
These formulae fit reasonably well the results obtained
by evolving the soft terms from $M_{mess} = 10^{13}$ GeV down to
$\mu = 1$ TeV with the code SUSPECT~\cite{suspect}.
For the third generation sfermion masses, most notably for $m^2_{Q_3}$
and $m^2_{U^c_3}$, the Yukawa couplings contribute sizeably
to the running and the above formulae do not apply. The Higgs
and neutralino/chargino spectrum also depend on $\tan \beta$
and on the values of the $\mu$ and $B \mu$ parameters,
which are determined from the requirement of proper radiative
electroweak symmetry breaking. As for the lightest neutralino,
Eq.~(\ref{eq:M_chi0_1}) implies that $M_{\tilde \chi^0_1} < m_{3/2}$
as long as $M_1 \lesssim 2 m_{3/2}$,
a condition which is unlikely to be violated
if $M_1$ is of gravitational origin, and we can therefore
safely assume that the lightest neutralino is the LSP.
The gravitino is then the NLSP, and its late decays into $\tilde \chi^0_1 \gamma$
tend to spoil the successful predictions of Big Bang nucleosynthesis if it is
abundantly produced after inflation.
This is the well-known gravitino problem~\cite{gravitino_problem},
and it is especially severe for a gravitino mass in the few $10$~GeV range,
as in our scenario. We are therefore led to assume a low reheating temperature
in order to reduce the gravitino abundance, typically
$T_R \lesssim (10^5 -10^6)$ GeV, which strongly disfavor baryogenesis
mechanisms occuring at very high temperatures, such as (non-resonant)
thermal leptogenesis.

While the lightness of $\tilde \chi^0_1$ is a welcome feature from the
point of view of distinguishing the present scenario from other
supersymmetric models (for recent studies of the collider signatures
of a light neutralino, see e.g. Refs.~\cite{BBCPR03,BFPS08}), it might be
a problem for cosmology. Indeed, a neutralino with a mass below,
say, $50$ GeV will generally overclose the universe, unless some
annihilation processes are very efficient~\cite{HP02,BFS02,BBCPR03}:
(i) the annihilation into
$\tau^+ \tau^-$ and $b \bar b$
via s-channel exchange of the CP-odd Higgs boson $A$, or
(ii) the annihilation into a fermion-antifermion pair via t- and u-channel
exchange of a light sfermion.
The process (i) can bring the relic neutralino abundance
down to the observed dark matter level (namely,
$\Omega_{DM} h^2 = 0.1099 \pm 0.0062$~\cite{WMAP})
if $A$ is light, $\tan \beta$ is large
and $\tilde \chi^0_1$ contains a sizeable higgsino component
(which requires $|\mu| \sim 100$ GeV). More precisely,
$\tilde \chi^0_1$ can be as light as $6$ GeV for $M_A \sim 90$ GeV
and $\tan \beta > 30$~\cite{BFS02,BBCPR03}, in the anti-decoupling
regime for the lightest Higgs boson $h$. The process (ii) is more
efficient for light sleptons ($\tilde l_R$) and large values of
$\tan \beta$. In particular, in the large $m_A$ region where
the process (i) is not relevant, $\tilde \chi^0_1$ can be as light
as $18$ GeV without exceeding the observed dark matter density
if $m_{\tilde \tau_1}$ is close to its experimental bound of $86$ GeV
and $\tan \beta \sim 50$~\cite{HP02,BBCPR03}.
Note that experimental limits on superpartner masses and rare processes
have been imposed in deriving these bounds.

We were not able to find values of $M_{GM}$, $N_m$ and $\tan \beta$
leading to a light $A$ boson (say, $M_A \leq 120$ GeV); hence
we must considerer $M_{\tilde \chi^0_1} > 18$ GeV in order
to comply with the dark matter constraint. In Table~\ref{tab:spectra},
we display 6 representative spectra with
$20\, \mbox{GeV} \leq M_{\tilde \chi^0_1} \leq 45\, \mbox{GeV}$
and light $\tilde l_R$ masses
(apart from model 1), corresponding to different numbers and types
of messengers, and different values of $M_{GM}$ and $\tan \beta$.
The superpartner masses were obtained by running the soft terms
from $M_{mess} = 10^{13}$ GeV down to low energy with the code
SUSPECT. Apart from $M_1$, which is taken as an input,
the unknown subdominant gravitational contributions to the soft terms
have not been included (we shall comment on this later).
As is customary, $\tilde f_1$ and $\tilde f_2$ refer to the lighter
and heavier $\tilde f$ mass eigenstates;
for the first two generations of sfermions, they practically coincide
with $\tilde f_R$ and $\tilde f_L$. We also indicated in Table~\ref{tab:spectra}
the bino and down higgsino components of the lightest neutralino,
in the notation $\tilde \chi^0_1 = Z_{11} \tilde B + Z_{12} \tilde W^3
+ Z_{13} \tilde H^0_d + Z_{14} \tilde H^0_u$.

\begin{table}
\vskip -1cm
\begin{center}
\begin{tabular}{|c|c|c|c|c|c|c|c|}
\hline model & 1 & 2 & 3 & 3 bis & 4 & 5 & 6 \\ 
\hline \hline
$N_{(5, \bar 5)}$                & $1$ & $6$ & $0$ & $0$
 & $0$ & $1$ & $3$ \\ 
$N_{(10, \overline{10})}$ & $0$ & $0$ & $1$ & $1$
 & $4$ & $1$ & $1$ \\  
$M_{GM}$ & $1000$ & $200$ & $300$ & $300$
 & $110$ & $220$ & $160$ \\  
$M_1$ & $50$ & $50$ & $50$ & $85$
 & $80$ & $85$ & $85$ \\
$\tan \beta$ & $30$ & $24$ & $15$ & $15$
 & $9$ & $15$ & $15$ \\ 
$\mbox{sign} (\mu)$ & $+$ & $+$ & $+$ & $+$
 & $+$ & $+$ & $+$ \\ 
\hline
$h$          & $114.7$ & $115.0$ & $115.2$ & $115.2$
 & $116.5$ & $114.6$ & $114.8$ \\ 
$A$          & $779.2$ & $645.4$ & $892.2$ & $892.4$
 & $1015$ & $735.8$ & $662.7$ \\ 
$H^0$     & $779.2$ & $645.5$ & $892.4$ & $892.6$
 & $1015$ & $735.9$ & $662.8$ \\ 
$H^\pm$ & $783.3$ & $650.3$ & $895.7$ & $895.9$
 & $1018$ & $740.1$ & $667.5$ \\ 
\hline
$\tilde \chi^\pm_1$ & $259.4$ & $305.0$ & $560.2$ & $560.3$
 & $676.7$ & $408.0$ & $223.9$ \\  
$\tilde \chi^\pm_2$ & $747.8$ & $636.8$ & $693.9$ & $694.0$
 & $970.4$ & $590.4$ & $597.5$ \\ 
\hline
$\tilde \chi^0_1$ & $24.5$ & $23.5$ & $23.2$ & $42.9$
 & $38.1$ & $43.0$ & $42.9$ \\ 
$\tilde \chi^0_2$ & $259.4$ & $305.0$ & $560.1$ & $560.3$
 & $677.1$ & $408.0$ & $223.9$ \\ 
$\tilde \chi^0_3$ & $743.3$ & $629.8$ & $596.9$ & $597.1$
 & $691.0$ & $570.8$ & $589.2$ \\ 
$\tilde \chi^0_4$ & $745.7$ & $634.7$ & $693.8$ & $693.9$
 & $970.4$ & $590.4$ & $596.3$ \\ 
\hline
$|Z_{11}|$ & $0.9982$ & $0.9975$ & $0.9971$ & $0.9971$
 & $0.9978$ & $0.9968$ & $0.9969$ \\ 
$|Z_{13}|$ & $0.0599$ & $0.0708$ & $0.0750$ & $0.0755$
 & $0.0648$ & $0.0792$ & $0.0772$ \\ 
\hline
$\tilde g$ & $1064$ & $1207$ & $1097$ & $1097$
 & $1527$ & $1028$ & $1063$ \\ 
\hline
$\tilde t_1$                    & $984.6$ & $927.3$ & $861.7$ & $861.6$
 & $1080$ & $795.7$ & $809.5$ \\ 
$\tilde t_2$                    & $1156$ & $1074$ & $1240$ & $1240$
 & $1468$ & $1058$ & $1002$ \\ 
$\tilde u_1, \tilde c_1$ & $1195$ & $1087$ & $1135$ & $1135$
 & $1361$ & $1006$ & $987.9$ \\ 
$\tilde u_2, \tilde c_2$ & $1240$ & $1115$ & $1327$ & $1327$
 & $1555$ & $1118$ & $1043$ \\
$\tilde b_1$                   & $1128$ & $1040$ & $1123$ & $1123$
 & $1356$ & $995.4$ & $966.2$ \\ 
$\tilde b_2$                   & $1169$ & $1079$ & $1224$ & $1224$
 & $1451$ & $1038$ & $987.1$ \\ 
$\tilde d_1, \tilde s_1$ & $1184$ & $1085$ & $1134$ & $1134$
 & $1360$ & $1005$ & $987.1$ \\ 
$\tilde d_2, \tilde s_2$ & $1243$ & $1117$ & $1329$ & $1329$
 & $1557$ & $1121$ & $1046$ \\ 
\hline
$\tilde \tau_1$                    & $242.2$ & $99.0$ & $86.3$ & $89.3$
 & $87.0$ & $96.7$ & $95.2$ \\ 
$\tilde \tau_2$                    & $420.3$ & $289.4$ & $696.2$ & $696.3$
 & $753.1$ & $498.6$ & $349.8$ \\ 
$\tilde e_1, \tilde \mu_1$ & $294.4$ & $150.6$ & $131.5$ & $133.6$
 & $105.4$ & $123.6$ & $117.4$ \\ 
$\tilde e_2, \tilde \mu_2$ & $413.4$ & $275.1$ & $699.1$ & $699.2$
 & $754.1$ & $500.1$ & $348.5$ \\ 
$\tilde \nu_{\tau}$              & $396.6$ & $260.5$ & $691.4$ & $691.5$
 & $749.0$ & $491.4$ & $337.6$ \\ 
$\tilde \nu_e, \tilde \nu_\mu$ & $405.8$ & $263.6$ & $694.8$ & $694.9$
 & $750.1$ & $493.9$ & $339.5$ \\ 
\hline \hline
$\Omega_{\tilde \chi^0_1} h^2$ & $6.40$ & $0.428$ & $0.279$ & $0.122$
 & $0.124$ & $0.118$ & $0.116$ \\
\hline
\end{tabular}
\end{center}
\caption{Supersymmetric mass spectra obtained by running the soft terms
from $M_{mess} = 10^{13}$ GeV down to low energy with the code SUSPECT
(all masses in GeV).}
\label{tab:spectra}
\end{table}

Let us now comment on these spectra. In the case of messengers
in $(5, \bar 5)$ representations, taking into account the LEP lower bound
on the lightest Higgs boson mass ($m_h \geq 114.4$~GeV) and
the experimental limits on the superpartner masses generally
leads to relatively heavy $\tilde l_R$ (see model 1), although
larger values of $\tan \beta$ yield a lighter $\tilde \tau_1$
(for instance, shifting $\tan \beta$ from $30$ to $50$ in
model 1 gives $m_{\tilde \tau_1} = 150$~GeV). However, one can
accommodate a lighter $\tilde \tau_1$ if one assumes a large number
of messengers, as exemplified by model 2. Light sleptons
are more easily obtained with messengers in $(10, \overline{10})$
representations (models 3/3bis and 4), or in both $(5, \bar 5)$
and ($10$, $\overline{10}$) representations (models 5 and 6).
Note that both $m_{\tilde \tau_1}$ and $m_{\tilde \mu_1, \tilde e_1}$
are close to their experimental limits in model 4.
Apart from the mass of the lightest neutralino (and to a lesser
extent of $\tilde l_R$), the low-energy spectrum very weakly
depends on the actual value of $M_1$ (compare models 3 and 3bis,
which only differ by the value of $M_1$).
In the last column of Table~\ref{tab:spectra}, we give the relic
density of $\tilde \chi^0_1$ computed by the code
micrOMEGAs~\cite{Micromegas,Micromegas_MSSM}.
One can see that, for $M_{\tilde \chi^0_1} \sim (20 - 25)$ GeV,
$\Omega_{\tilde \chi^0_1} h^2$ lies above the observed
dark matter density, even though $\tilde l_R$ are light (models 1 to 3);
this can be traced back to the small higgsino admixture of $\tilde \chi^0_1$,
which suppresses the $Z$ boson exchange contribution~\cite{BBCPR03}.
Larger values of $M_{\tilde \chi^0_1}$ enable the relic density to fall
in the $2 \sigma$ WMAP range (models 3bis to 6).

We conclude that the scenario of supersymmetry breaking
considered in this paper can provide supersymmetric models
with a light neutralino ($M_{\tilde \chi^0_1} \sim 40$ GeV)
accounting for the dark matter of the universe.
We have checked that the models of Table~\ref{tab:spectra}
are consistent with the negative results from direct dark matter detection
experiments such as CDMS~\cite{CDMS} and XENON~\cite{XENON}.
Since the spin-independent (spin-dependent) neutralino-nucleon
cross section is dominated by Higgs boson and squark exchange
diagrams ($Z$ boson and squark exchange diagrams), it is expected
to be rather small in our scenario, in which squarks are heavy
and the neutralino is mostly a bino. This is confirmed by a
numerical computation with MicrOMEGAs, which gives typical
values of $(10^{-46} - 10^{-45})$ cm$^2$ for the spin-independent
cross-section, and of $(10^{-46} - 10^{-45})$ cm$^2$ for the
spin-dependent cross-section.

Let us add for completeness that models 1 to 3 can be made consistent
with the observed dark matter density by assuming a small amount
of R-parity violation~\cite{Dreiner}. In fact, in the presence of $R$-parity
violation, nothing prevents us from considering even smaller
neutralino masses by lowering\footnote{Assuming
$M_1 \sim m_{3/2}$, one can reach $M_{\tilde \chi^0_1} \sim 5$ GeV
by choosing $m_{3/2} \sim 10$ GeV. We refrain from considering much
lower values of $m_{3/2}$, which would render the generation
of $\mu \sim M_{GM}$ less natural. However, we note that in recent
models of moduli stabilization~\cite{stab1,stab2}, gravity (moduli)
contributions to gaugino masses are typically smaller than
$m_{3/2}$ by one order of magnitude.} $m_{3/2}$.

Some comments are in order regarding the subdominant
supergravity contributions to the soft terms
and their effects in flavor physics. First of all, these contributions
will shift the values of the soft terms at $M_{mess}$ by
a small amount and correspondingly affect the spectra
presented in Table~\ref{tab:spectra}. Since supergravity contributions
are parametrically suppressed with respect to gauge contributions
by a factor $m_{3/2} / (N_m M_{GM})$ for gaugino masses,
and by a factor $m_{3/2} / (\sqrt{N_m} M_{GM})$ for scalar masses,
we do not expect them to change the qualitative features of the
spectra\footnote{For values of $m_{3/2}$ as large as
$80-85$ GeV, however, the supergravity contribution to the
$\tilde l_R$ masses is expected to be comparable to the GMSB one.
In this case the parameters of the models in Table~\ref{tab:spectra}
must be adjusted in order to keep the sleptons sufficiently light.}.
Also, the gravity-mediated $A$-terms are suppressed by the
small gravitino mass, and they should not affect the sfermion
masses in a significant way. The most noticeable consequence
of the supergravity contributions is actually to introduce flavor
violation in the sfermion sector at the messenger scale:
\begin{equation}
  (M^2_\chi)_{ij}\ =\ m^2_\chi\, \delta_{ij}\, +\, (\lambda_\chi)_{ij}\, m^2_{3/2}
  \qquad  (\chi = Q, U^c, D^c, L, E^c)\ ,
\end{equation}
where $m^2_{\chi}\, \delta_{ij}$ is the flavor-universal gauge-mediated 
contribution, and the coefficients $(\lambda_\chi)_{ij}$ are at most
of order one. As is well known, flavor-violating processes
are controlled by the mass insertion parameters
(here for the down squark sector):
\begin{equation}
  ( \delta^d_{LL})_{ij}\, \equiv\, \frac{(M^2_Q)_{ij}}{\bar m^2_{\tilde d}}\, , \quad
  (\delta^d_{RR})_{ij}\, \equiv\, \frac{(M^2_{D^c})_{ij}}{\bar m^2_{\tilde d}}\, , \quad
  (\delta^d_{LR})_{ij}\, \equiv\, \frac{(A_d)_{ij} v_d}{\bar m^2_{\tilde d}}
  \qquad (i \neq j)\ ,
\label{eq:deltas}
\end{equation}
where $(M^2_Q)_{ij}$, $(M^2_{D^c})_{ij}$ and $(A_d)_{ij} v_d$ are
the off-diagonal entries of the soft scalar mass matrices renormalized
at low energy and expressed in the basis of down quark mass eigenstates,
and $\bar m_{\tilde d}$ is an average down squark mass.

Neglecting the RG-induced flavor non-universalities,
which are suppressed by a loop factor and by small CKM angles,
the mass insertion parameters $(\delta^d_{MM})_{ij}$ ($M=L,R$)
arising from the non-universal supergravity contributions are
suppressed by a factor $m^2_{3/2} / \bar m^2_{\tilde d}\, $, and possibly
also by small coefficients $(\lambda_{Q,D^c})_{ij}$.
For $m_{3/2} = 85$ GeV and $\bar m_{\tilde d} \sim 1$ TeV
as in the spectra displayed in Table~\ref{tab:spectra}, we find
$(\delta^d_{LL})_{ij} \sim 7 \times 10^{-3}\, (\lambda_Q)_{ij}$ and
$(\delta^d_{RR})_{ij} \sim 7 \times 10^{-3}\, (\lambda_{D^c})_{ij}$,
which is sufficient to cope with all experimental constraints
(in the presence of large CP-violating phases, however, $\epsilon_K$
further requires $\sqrt{(\lambda_Q)_{12} (\lambda_{D^c})_{12}} \lesssim 0.04$,
see e.g. Ref.~\cite{MVV07}). As for the $(\delta^d_{LR})_{ij}$,
they are typically suppressed by $m_{3/2} m_{d_i} / \bar m^2_{\tilde d}$
and are therefore harmless.

The situation is much more problematic in the slepton sector,
where processes such as $\mu \rightarrow e \gamma$ and
$\tau \rightarrow \mu \gamma$ put strong constraints on the
$(\delta^e_{MN})_{ij}$, $M,N=L,R$ (see e.g. Ref.~\cite{MS02}).
Indeed, the leptonic $\delta$'s are less suppressed than the
hadronic ones, due to the smallness of the slepton masses:
for $m_{3/2} = 50$ GeV and $m_{L_i} \sim 500$ GeV,
$m_{E^c_i} \sim 100$ GeV, one e.g. finds
$(\delta^e_{LL})_{ij} \sim 10^{-2} (\lambda_L)_{ij}$ and
$(\delta^e_{RR})_{ij} \sim 0.3\, (\lambda_{E^c})_{ij}$.
To cope with the experimental
constraints, which are particularly severe in the presence
of a light neutralino and of light sleptons, we need to assume
close to universal supergravity contributions to slepton
soft masses, perhaps due to some flavor symmetry responsible
for the Yukawa hierarchies. Possible other sources of lepton
flavor violation, e.g. radiative corrections induced by heavy states,
should also be suppressed. Let us stress that the same problem
is likely to be present in any light neutralino scenario in which
the neutralino annihilation dominantly proceeds through
slepton exchange.
Alternatively, in models where the relic density of $\tilde \chi^0_1$
is controlled by a small amount of R-parity violation, all sleptons
can be relatively heavy as in model 1, thus weakening the
constraints from the non-observation of lepton flavor violating
processes.

Throughout this paper, we assumed that the non-renormalizable
operator $\Phi \Sigma^2 \tilde \Phi / M_P$ is absent from the
superpotential and that $M_1$ is purely of gravitational origin.
Let us mention for completeness the alternative possibility
that this operator is present and gives the dominant contribution
to $M_1$. In this case, the lightest neutralino mass
is no longer tied up with the mass of the gravitino, which can be
the LSP as in standard gauge mediation. This makes it possible
to solve the lepton flavor problem by taking $m_{3/2} \lesssim 10$ GeV
and considering a model with relatively heavy $\tilde l_R$. Such a scenario
is still characterized by a light neutralino, but it is no longer the LSP,
and the dark matter abundance is no longer predicted in terms
of parameters accessible at high-energy colliders.
Furthermore, the superpartner spectrum depends
on an additional parameter, the coefficient of the non-renormalizable
operator $\Phi \Sigma^2 \tilde \Phi / M_P$.

\section{Conclusions}
\label{sec:conclusions}

In this paper, we have shown that models in which supersymmetry
breaking is predominantly transmitted by gauge interactions lead
to a light neutralino if the messenger mass matrix is oriented with
the hypercharge generator, $M \sim v Y$. This arises naturally
if the main contribution to messenger masses comes from a coupling
to the adjoint Higgs field of an underlying $SU(5)$ theory.
In this case, the bino receives its mass from gravity mediation, leading
to a light neutralino which is then the LSP.
While from a model building perspective the gravitino, hence the
neutralino, could be much lighter, we considered a typical neutralino
mass in the $(20-45)$ GeV range and worked out the corresponding
low-energy superpartner spectrum. We noticed that,
in the case of $(10, \overline{10})$ messengers or of a large number
of $(5, \bar 5)$ messengers, the scalar partners of the right-handed leptons
are much lighter than the other sfermions,
making it possible for a neutralino with a mass around $40$ GeV
to be a viable dark matter candidate. However, such a SUSY spectrum
also creates potential FCNC problems in the lepton sector, which asks
for a high degree of universality or alignment in slepton masses.

In the hybrid models of supersymmetry breaking considered in this paper,
the gravity-mediated contributions, although subdominant,
are essential in generating the $\mu$ and $B \mu$ terms
through Planck-suppressed operators. 
We studied the case where the SUSY breaking sector
is provided by the ISS model and found that, as expected,
messenger loops induce a breaking of the R-symmetry in the ISS vacuum.
The associated meson vev's
happen to be of the appropriate size for generating
the $B \mu$ term needed for electroweak symmetry breaking. 
We stress that this mechanism also works for more general messenger
mass matrices than the one studied in this paper, in particular in the simpler
case of $SU(5)$ symmetric messenger masses.

While the vanishing of the GMSB contribution to the bino mass
is a simple consequence of the messenger mass matrix~(\ref{i5bis})
and of the embedding of the hypercharge into a simple gauge group,
the other features of the superpartner spectrum depend on the
representation of the messengers, in contrast to minimal
gauge mediation. For example, the gluino to wino mass ratio is
$|M_3/M_2| = 3 \alpha_3 / 2 \alpha_2$ for $(5, {\bar 5})$ messengers
and  $|M_3/M_2| = 7 \alpha_3 / 12 \alpha_2$
for $(10, \overline{10})$ messengers. The experimental evidence
for one of these mass ratios at the LHC, together with the discovery
of a light neutralino LSP, would be a clear signature of the hybrid
models of supersymmetry breaking studied in this paper.
In most high-energy scenarios, gaugino masses are assumed
to be universal, leading to the hierarchy
$M_1 : M_2 : M_3 = \alpha_1 : \alpha_2 : \alpha_3$ at
low energy. The possibility that non-universal gaugino masses
be related to the lightness of the neutralino LSP by an underlying
GUT structure appears to be appealing and deserves further investigation.


\section*{Acknowledgments}
{We thank Genevi\`eve B\'elanger, Marco Cirelli, Tony Gherghetta,
Yann Mambrini, Mariano Quiros, Alberto Romagnoni and Carlos Savoy
for useful discussions and comments.
We are grateful to Yann Mambrini for providing us with an improved version
of the code SUSPECT. This work has been supported in part by the ANR grants
ANR-05-BLAN-0079-02, ANR-05-BLAN-0193-02, ANR-05-JCJC-0023, the RTN
contracts MRTN-CT-2004-005104 and MRTN-CT-2004-503369, the CNRS PICS
\#~2530 and 3747 and the European Union Excellence Grant MEXT-CT-2003-509661.}


\renewcommand{\theequation}{A.\arabic{equation}}
\setcounter{equation}{0}  


\begin{appendix}

\section{Gauge contributions to the MSSM gaugino
and scalar masses}
\label{app1}

In this appendix, we compute the gauge-mediated contributions
to the MSSM soft terms in the scenario with a GUT-induced
messenger mass splitting considered in this paper.
We use the method of Ref.~\cite{gr2}, appropriately generalized
to the case of several types of messengers with different masses.

\subsection{General formulae}

The gauge-mediated contributions to gaugino masses
are encoded in the running of the gauge couplings~\cite{gr2}:
\begin{equation}
\frac{1}{g_a^2(\mu)} \ = \
\frac{1}{g_a^2(\Lambda_{UV})} \ - \ \frac{b_a}{8 \pi^2}\,
\ln \left(\frac{\Lambda_{UV}}{\mu}\right)
+\, \sum_i\, \frac{2 T_a (R_i)}{8 \pi^2}\,
\ln \left(\frac{\Lambda_{UV}}{\mbox M_i}\right) \ , \label{mssm03}
\end{equation}
where $b_a = 3 C_2 (G_a) - \sum_R T_a (R)$ is the beta function
coefficient of the gauge group factor $G_a$, and the sum runs over
several types of messengers $(\phi_i, {\tilde \phi_i})$ with masses $M_i$
($\mu < M_i < \Lambda_{UV}$) belonging to the SM gauge representations
$R_i$. $T_a(R_i)$ is the Dynkin index of the representation $R_i$,
normalized to $1/2$ for fundamental representations of $SU(N)$.
For $U(1)$, we use the $SU(5)$
normalization $\alpha_1 = \frac{5}{3}\, \alpha_Y$; correspondingly,
$T_1 (R_i)$ should be understood as $3 Y^2_i / 5$, where the hypercharge
$Y$ is defined by $Y = Q - T_3$ (so that $Y_Q= 1/6$, $Y_{U^c}= -2/3$,
$Y_{D^c}= 1/3$, $Y_L= - 1/2$ and $Y_{E^c}= 1$). The one-loop gaugino
masses are then given by~\cite{gr2}
\begin{equation}
M_{a} (\mu) \ = \ \frac{\alpha_a(\mu)}{4\pi}\ \sum_i\, 2 T_a (R_i)\,
\frac{\partial \ln (\det M_i )}{\partial \ln X}
\left. \frac{F_X}{X}\, \right|_{X=X_0}\ .
\label{mssm04}
\end{equation}

The gauge-mediated contributions to scalar masses are encoded
in the wave-function renormalization of the MSSM chiral
superfields $\chi$~\cite{gr2}:
\begin{equation}
Z_\chi (\mu) \ = \ Z_\chi (\Lambda_{UV})\ \prod_a\,
  \left(\frac{\alpha_a(\Lambda_{UV})}{\alpha_a(M_2)}\right)^{\frac{2C^a_\chi}{b_{a, 2}}}
  \left( \frac{\alpha_a (M_2)}{\alpha_a (M_1)} \right)^{\frac{2C^a_\chi}{b_{a, 1}}}
  \left( \frac{\alpha_a (M_1)}{\alpha_a (\mu)} \right)^{\frac{2C^a_\chi}{b_a}}  ,
\label{eq:Z_Phi}
\end{equation}
where $\mu < M_1 < M_2 < \Lambda_{UV}$,
$b_{a,1} \equiv b_a - 2 N_1 T_a(R_1)$,
$b_{a,2} \equiv b_{a,1} - 2 N_2 T_a(R_2)$,
and $C^a_\chi$ are the quadratic Casimir coefficients
for the superfield $\chi$, normalized to $C (N) = (N^2-1)/2N$ 
for the fundamental representation of $SU(N)$ and to
$C^1_\chi = 3 Y^2_\chi / 5$ for $U(1)$. 
In Eq.~(\ref{eq:Z_Phi}), we
considered for simplicity only 2 types of messengers, characterized
by their masses $M_{1,2}$ (which should not be confused with
the bino and wino masses), SM gauge representations $R_{1,2}$
and multiplicities $N_{1,2}$. Following Ref.~\cite{gr2}, we obtain for
the soft mass parameter $m^2_\chi$:
\begin{eqnarray}
m^2_\chi & = & 2 \sum_a C^a_\chi \left( \frac{\alpha_a (\mu)}{4\pi} \right)^2
  \biggl\{ \left[\, 2 N_2 T_a(R_2)\xi^2_{a,2}
  +\, \frac{(2 N_2 T_a(R_2))^2}{b_{a,1}}\, (\xi_{a,1}^2 -\xi_{a,2}^2)\, \right]
  \left| \frac{\partial\ \mbox{ln} M_2}{\partial \mbox{ln} X} \right|^2  \nonumber \\
&& +\ 2 N_1 T_a(R_1) \xi_{a,1}^2 \left| \frac{\partial \ln M_1 }{\partial \ln X} \right|^2
  +\, \frac{1 -\xi_{a,1}^2}{ b_a} \left| \frac{\partial \ln (\det M)}{\partial \ln X} \right|^2\,
  \biggr\} \left. \left| \frac{F_X}{X} \right|^2\, \right|_{X=X_0}\, ,
\label{a6}
\end{eqnarray}
where $\xi_{a,i} \equiv \frac{\alpha_a (M_i)}{\alpha_a (\mu)}$ ($i=1,2$)
and $\det M = M_1^{N_1} M_2^{N_2}$. In Eq.~(\ref{a6}), the first term
in square brackets contains the contribution of the messengers of mass
$M_2$ renormalized at the scale $M_1$,
the second term represents the contribution of the messengers of mass $M_1$,
and the third term the running from the messenger scale $M_1$
down to the low-energy scale $\mu$.

\subsection{($5$, $\bar 5$) and ($10$, $\overline{10}$) messengers
with GUT-induced mass splitting}

We are now in a position to evaluate the MSSM gaugino and scalar
masses induced by $N_m$ ($5, \bar 5$) messenger pairs with a mass
matrix $M(X)$ given by Eq.~(\ref{i5bis}). Inside each pair,
the $SU(3)_C$ triplets have a mass $2 \lambda_{\Sigma} v$,
while the $SU(2)_L$ doublets have a mass $- 3 \lambda_{\Sigma} v$
(we omit the contribution of $X_0 \neq 0$, which as discussed
in Section~\ref{subsec:coupling} turns out to be negligible).
Applying Eq.~(\ref{mssm04}), we obtain for the one-loop gaugino masses:
\begin{equation}
M_3\ =\ \frac{1}{2}\, N_m\, \frac{\alpha_3}{4 \pi}\, \frac{\lambda_X F_X}
  {\lambda_{\Sigma} v}\ , \qquad M_2\ =\ - \frac{1}{3}\, N_m\,
  \frac{\alpha_2}{4 \pi}\, \frac{\lambda_X F_X}{\lambda_{\Sigma} v}\ ,
  \qquad M_1\ =\ 0 \ .
\label{eq:Ma_5}
\end{equation}
In computing scalar masses, we neglect for simplicity the running
of the gauge couplings between different messenger scales,
which amounts
to set $\alpha_a (M_1) = \alpha_a(M_2) \equiv \alpha_a (M_{mess})$
in Eq.~(\ref{a6}), where $M_{mess}$ is an average messenger mass.
Summing up all gauge contributions, we can cast the scalar masses
in the form
\begin{equation}
m^2_\chi (M_{mess})\ =\  N_m \sum_a d^a_\chi
  \left( \frac{\alpha_a}{4 \pi} \right)^2
  \left| \frac{\lambda_X F_X}{\lambda_\Sigma v} \right|^2  ,
\label{eq:m2_Phi}
\end{equation}
where $\alpha_a = \alpha_a (M_{mess})$ and the coefficients $d^a_\chi$
are given in the following table:

$$
\begin{array}{|c|c|c|c|c|}
\hline
 d^a_\chi & SU(3)_C& SU(2)_L& U(1) \\
\hline
 Q & 2/3 &  1/6 & 1/180 \\
\hline
 U^c & 2/3 &  0 & 4/45 \\
\hline
 D^c & 2/3 &  0 & 1/45 \\
\hline
 L & 0 &  1/6 & 1/20 \\
\hline
E^c & 0 & 0 & 1/5 \\
\hline
 H_u, H_d & 0 &  1/6 & 1/20 \\
\hline
\end{array}
$$

Consider now $N_m$ $(10, \overline{10})$ messenger pairs.
Inside each pair, the ($\phi_{3, 2,+1/6}$, $\tilde \phi_{\bar 3, 2,-1/6}$)
fields have a mass $\lambda_\Sigma v$,
($\phi_{\bar 3, 1, - 2/3}$, $\tilde \phi_{\bar 3, 1, +2/3}$)
have a mass $- 4 \lambda_\Sigma v$,
and ($\phi_{1, 1, +1}$, $\tilde \phi_{1, 1, -1}$)
have a mass $6 \lambda_\Sigma v$. Then the
gaugino masses are given by
\begin{equation}
M_3\ =\ \frac{7}{4}\, N_m\, \frac{\alpha_3}{4 \pi}\, \frac{\lambda_X F_X}
  {\lambda_{\Sigma} v}\ , \qquad M_2\ =\ 3\, N_m\,
  \frac{\alpha_2}{4 \pi}\, \frac{\lambda_X F_X}{\lambda_{\Sigma} v}\ ,
  \qquad M_1\ =\ 0 \ ,
\label{eq:Ma_10}
\end{equation}
and the scalar masses by Eq.~(\ref{eq:m2_Phi}), with
coefficients $d^a_\chi$ given by: 
$$
\begin{array}{|c|c|c|c|c|}
\hline
 d^a_\chi& SU(3)_C& SU(2)_L& U(1) \\
\hline
 Q & 11/2 &  9/2 & 1/90 \\
\hline
 U^c & 11/2 &  0 & 8/45 \\
\hline
 D^c & 11/2 &  0 & 2/45 \\
\hline
 L & 0 &  9/2 & 1/10 \\
\hline
E^c & 0 & 0 & 2/5 \\
\hline
H_u, H_d & 0 &  9/2 & 1/10 \\
\hline
\end{array}
$$


\renewcommand{\theequation}{B.\arabic{equation}}
\setcounter{equation}{0}  

\section{Quantum corrections and metastability of the vacuum}
\label{app2}

\subsection{Tree-level vacuum structure}

We are searching for the minima of the scalar potential
\begin{equation}
V\ =\ |F_X^a|^2 + |F_X^b|^2 + |F_q|^2 + |F_{\tilde{q}}|^2+
|F_\phi|^2 + |F_{\tilde{\phi}}|^2+ |F_\Sigma|^2 \ ,
\label{tree1}
\end{equation}
where
\begin{eqnarray}
 |F_X^a|^2 &=& \displaystyle\sum_{i=1}^N  \left| h q^i_a \tilde{q}_{i}^a  -h f^2 +
 \lambda_{X,i}^i \phi \tilde{\phi}\right|^2 \ ,
\nonumber \\
 |F_X^b|^2 &=& \displaystyle\sum_{(i,j) \notin \{i=j=1 \ldots N \} } \left| -h f^2 \delta^i_j+
 \lambda_{X,j}^i \phi \tilde{\phi} \right|^2 \ ,
\nonumber \\
|F_q|^2 &=& \displaystyle\sum_{a,i = 1 \ldots N}  \left| h
X_i^{j}\tilde{q}_{j}^a \right|^2 \ ,
\nonumber \\
|F_{\tilde{q}}|^2 &=&  \displaystyle\sum_{a,j = 1 \ldots N} \left|h
q^i_a X_i^{j} \right|^2 \ ,
\\
|F_{\phi}|^2 &=& \left|  (\lambda_X X + \lambda_{\Sigma} \Sigma)
\tilde{\phi} \right|^2 \ ,
\nonumber \\
|F_{\tilde{\phi}}|^2 &=&\left|  \phi (\lambda_X X +
\lambda_{\Sigma} \Sigma)  \right|^2 \ ,
\nonumber \\
|F_\Sigma|^2 & = &  \left|\lambda_\Sigma \phi \tilde{\phi} +
\frac{\partial W_{GUT}}{\partial \Sigma}\right|^2 \ .
\nonumber
\end{eqnarray}
We choose a basis in which $q^i_a \tilde{q}_{j}^a $ is a rank
N diagonal matrix:
\begin{equation}
\left(
\begin{array}{ccc|cc}
q_1 \tilde q_1  & 0 & 0&0&0 \\
0& \cdots & 0 &0&0\\
0 & 0 &q_N \tilde q_N  &0&0 \\ \hline
0 &0 &0 &0&0\\
0 &0 &0 &0&0
\end{array}
\right)\ .
\end{equation}

The potential~(\ref{tree1}) does not contain the supergravity
contributions nor the corresponding soft terms, which are
expected to have a negligible impact in the present discussion.
$W_{GUT} (\Sigma)$ is the superpotential for the
$SU(5)$ adjoint Higgs field $\Sigma$, whose vev is responsible
for the spontaneous breaking of $SU(5)$.

We find that all the F-terms, except $F_X^b$, can be set to zero.
However, $F_{\phi}=F_{\tilde \phi}= 0$ has two types of solutions.
More precisely, for values of $X$ such that the matrix (acting on
$SU(5)$ gauge indices)
$\lambda_X X + \lambda_\Sigma \Sigma $ is
\begin{itemize}
\item invertible, then $\phi$ =$\tilde{\phi}=0$;
\item non invertible, then both $\phi$ and $\tilde{\phi}$ can have
a non zero vev.
\end{itemize}
Indeed, if $\lambda_X X + \lambda_{\Sigma} \sigma_i =0 $,
where $\sigma_i$ is an eigenvalue of $\Sigma$, the values of $\phi$
and $\tilde{\phi}$ are not fixed by the constraint $F_\phi = F_{\tilde \phi} = 0$.
The equation $F_\Sigma=0$ implies that they must be of the form
$\phi = (0, \ldots , 0, \phi^{\alpha_0}, 0, \ldots , 0)$ and
$\tilde \phi^T = (0, \ldots , 0, \tilde \phi_{\alpha_0}, 0, \ldots , 0)$.
Indeed, one has $F_{\Sigma,\beta}^\alpha
= f'(\Sigma)^\alpha_\beta - \rho\, \delta^\alpha_\beta
+ \lambda_\Sigma \phi^\alpha \tilde \phi_\beta =0$,
where $\alpha, \beta = 1 \ldots 5$ are $SU(5)$ indices
and $f (\Sigma)$ is defined by
$W_{GUT} (\Sigma) = f (\Sigma) - \rho\, \mbox{Tr} \Sigma$
(the specific form of the function $f$ is irrelevant here).
Working in a $SU(5)$ basis in which $\Sigma_\alpha^\beta$
is diagonal, one concludes that at most one component
in $\phi$ and $\tilde \phi$ can be nonzero, and it must be
the same component. As for $F_q$ and $F_{\tilde{q}} $,
they can always be fixed to zero by choosing the matrix
$X_i^j$ to be symmetric, with the vectors $q_a^i = \tilde q^a_i$
($a=1 \ldots N$), solutions of $h q^i_a \tilde{q}_{i}^a - h f^2
+ \lambda_{X,i}^i \phi \tilde \phi = 0$ (so as to satisfy the
constraint $|F_X^a|^2 = 0$), belonging to its kernel. Note that the value
of $X$ is not completely determined at this level. 

We have succeeded to set all F-terms but $F_X^b$  to zero
without completely fixing the value of $X$. For generic couplings
$\lambda^i_{X,j}$, it is still possible to arrange for the matrix
$\lambda_X X + \lambda_\Sigma \Sigma$ to have a zero eigenvalue,
in which case $\phi$ and $\tilde \phi$ can be nonzero. We can
minimize $|F_X^b|^2$ in both cases ($\phi \tilde \phi =0$ versus
$\phi \tilde \phi \neq 0$), which yields two types of
local supersymmetry-breaking minima:
\begin{itemize}
\item  $\phi \tilde{\phi}=0$, with the ISS energy
$V_0= (N_f-N) h^2 f^4$;
\item  $\phi \tilde{\phi} = \frac{-\displaystyle\sum_{i=N+1}^{N_f}
\lambda_{X,i}^i h f^2  }{\displaystyle\sum_{(i,j) \notin \{i=j=1 \ldots N \}}
\left|\lambda_{X,j}^i\right|^2} $ , with  $V_0=h^2 f^4
\left(N_f-N   - \frac{\left| \displaystyle\sum_{i=N+1}^{N_f}
\lambda_{X,i}^i\right|^2 }{ \displaystyle\sum_{(i,j) \notin \{i=j=1 \ldots N \}}
\left|\lambda_{X,j}^i\right|^2 }\right) .$
\end{itemize}
%

\subsection{Lifetime of the metastable vacuum}
\label{app2_lifetime}

Following Ref.~\cite{iss}, we evaluate the lifetime of the metastable
ISS vacuum in the triangle approximation. The decay rate is
proportional to
\begin{equation}
\exp \left( -\frac{(\Delta \phi)^4}{\Delta V} \right) , \quad  {\rm with}
\quad  \frac{\Delta V}{(\Delta \phi)^4}\ = \displaystyle
\sum_{(i,j) \notin \{i=j=1 \ldots N\}} |\lambda_{X,j}^i|^2\ \equiv\ \overline{\lambda^2}\ .
\end{equation}
In order for the metastable vacuum to be sufficiently long lived,
we require $\overline{\lambda^2} \lesssim 10^{-3}$. The individual
couplings $\lambda^i_{X,j}$ must then typically be of order $10^{-2}$,
except the ones corresponding to $i=j=1 \ldots N$, which can in principle
be larger. From Eq.~(\ref{i4}) we see that, for 
$\mbox{Tr}' \lambda \equiv {\sum_{i=N+1}^{N_f}} \lambda_{X,i}^i = 10^{-2}$,
$M_{GM} / m_{3/2} \sim 10$ corresponds to a messenger scale
$\lambda_\Sigma v \sim 10^{13}$ GeV, which in turn requires
$\lambda_\Sigma \sim 10^{-3}$.

\subsection{Quantum corrections to the scalar potential}

As explained in Ref.~\cite{iss}, the ISS model has a tree-level
flat direction along the $i,j = (N+1) \ldots N_f$ components of $X_i^j$.
In the absence of messengers, quantum corrections enforce
$\langle X \rangle = 0$. In this section, we add the contribution
of the messengers to the one-loop effective potential for $X$
and study its behaviour around
$\phi = \tilde{\phi}=0$. Our aim is to determine whether
the ISS vacuum remains metastable and long lived in our
scenario after quantum corrections have been included.

We parametrize the quantum fluctuations in the following way:
\begin{equation}
X\ =\ \left(
  \begin{array}{cc}
  \tilde{Y} & \delta Z^{\dagger} \\
  \delta \tilde{Z} & \tilde{X}
  \end{array}
    \right) , \qquad 
q\ =\ \left(
  f e^\theta + \delta \chi, \delta \rho
  \right) , \qquad
\tilde{q}\ =\ \left(
  \begin{array}{c}
  f e^{-\theta} + \delta \tilde \chi^\dagger\\
   \delta \tilde \rho^\dagger
  \end{array}
  \right) ,
\end{equation}
with $\tilde{X}= X_0 + \delta \hat{X}$ and $\tilde{Y} = Y_0 + \delta \hat {Y}$.
The only F-term from the ISS sector that is relevant for the computation
of the messenger contribution to the one-loop effective potential is
the one of $\tilde{X}$:
\begin{equation}
- F^\star_{\tilde{X}_{ff'}} =\
  h\, \mbox{Tr}_{N_c}(\delta \rho\, \delta \tilde{\rho}^{\dagger})_{f f'}
  - h f^2\, \delta_{f f'} + \lambda_{X, f f'}\, \delta \phi\, \delta \tilde{\phi} \ ,
\end{equation}
where $f, f' = (N+1) \ldots N_f$. The terms of the scalar potential
that contribute to the scalar messenger mass matrix are:
\begin{equation}
| h \mbox{Tr}_{N_c}(\delta \rho \delta \tilde{\rho}^{\dagger})_{f f'}
-h f^2 \delta_{f f'} +  \lambda_{X, f f'} \delta \phi
\delta \tilde{\phi}|^2 + |( \lambda_X X + m )\delta
\tilde{\phi}|^2 + |\delta \phi (\lambda_X X + m ) |^2 .
\end{equation}
Around the vacuum with zero messenger vev's, $\phi = \tilde{\phi}=0$,
there is no quadratic mixing between the ISS and messenger fields.
We can therefore compute separately the contributions of the ISS
and messenger sectors to the effective potential.

Let us first consider the messenger sector. With the notations
$\tilde{M}_I \equiv \lambda_X X + m_I$ (where the index $I$ refers
to different components of the messenger fields in definite SM
gauge representations, and $m_I = 6 \lambda_\Sigma Y_I v$),
$\mbox{Tr}' \lambda \equiv {\sum_{i=N+1}^{N_f}} \lambda_{X,i}^i$ and
$t \equiv h f^2\, \mbox{Tr}'\lambda$, the scalar mass matrix reads:
\begin{equation}
\left( \phi_I^{\dagger} \mbox{  } \tilde \phi_I^\dagger
  \mbox{  } \phi_I \mbox{  } \tilde\phi_I \right)
\left( \begin{array}{cccc}
  |\tilde{M}_I|^2 & & & -t^* \\
  & |\tilde{M}_I|^2 & -t^* & \\
  & -t & |\tilde{M}_I|^2 & \\
  -t & & &|\tilde{M}_I|^2 \ .
  \end{array} \right)
\left( \begin{array}{c}
  \phi_I \\ \tilde \phi_I \\ \phi_I^\dagger \\ \tilde \phi_I^\dagger
  \end{array} \right)\ .
\end{equation}
We then find the mass spectrum (which is non-tachyonic since
$|t| = |\lambda_X F_X| \ll \lambda^2_\Sigma v^2 \sim m^2_I$):
\begin{equation}
  m_{0,I}^2\ =\ |\tilde{M}_I|^2 \pm |t|\
  =\  |\lambda_X X + m_I|^2 \pm  h f^2 |\mbox{Tr}'\lambda| \ .
\end{equation}
The contribution of the messenger sector to the effective potential is then:
\begin{equation}
V_{\phi, \tilde{\phi}}^{(1)} \ = \ \frac{1}{64 \pi^2}\ \mbox{Str} M^4\,
  \ln \left( \frac{M^2}{\Lambda^2} \right)
  = \ \frac{2 N_m}{64 \pi^2} \left( 20 |t|^2 + 2 |t|^2
  \ln \left( \frac{\det \tilde{M}^{\dagger}\tilde{M}}{\Lambda^2}\right)\right) \ .
\end{equation}
As for the contribution of the ISS sector, it is given by~\cite{iss}:
\begin{equation}
  V_{ISS}^{(1)} \ = \ \frac{1}{64 \pi^2}\ 8\, h^4 f^2 (\ln 4 -1) N
  (N_f - N) |X_0|^2 \ ,
\end{equation}
where we have set $\tilde X = X_0\, \unit_{N_f-N}$,
$\tilde Y = Y_0\, \unit_N$, and we have omitted
a term proportional to $|Y_0|^2$, which is subleading with respect to the
tree-level potential for $Y_0$, $V_{ISS}^{(0)} (Y_0) = 2 N h^2 f^2 |Y_0|^2$
(by contrast, the term proportional to $|X_0|^2$ in $V_{ISS}^{(1)}$ is fully
relevant, since there is no tree-level potential for $X_0$).
To $V_{ISS}^{(0)} + V_{ISS}^{(1)}$, we add the linearized
field-dependent one-loop contribution of the messenger sector,
using the fact that $|t| \ll \lambda^2_\Sigma v^2$:
\begin{equation}
V_{\phi, \tilde{\phi}}^{(1)} \ = \  \frac{N_m |\mbox{Tr}' \lambda|^2 h^2 f^4 }
  {64 \pi^2} \left[ -\, \frac{35}{18 \lambda_\Sigma^2 v^2}\,
  (\lambda_X X)^2\, +\, \frac{10}{3 \lambda_\Sigma v}\,
  \lambda_X X + {\rm h.c.} \right] \ .
\end{equation}
As will become clear after minimization of the full one-loop
effective potential,
the quadratic term in $V_{\phi, \tilde{\phi}}^{(1)}$ is suppressed
with respect to the quadratic terms in $V_{ISS}$ by
$\langle X \rangle \ll \lambda_\Sigma v$, and can therefore be dropped.
Minimizing $V_{ISS}^{(0)} + V_{ISS}^{(1)} + V_{\phi, \tilde{\phi}}^{(1)}$,
one finds that the contribution of the messenger fields to the effective
potential destabilizes the tree-level ISS vacuum and creates
small tadpoles for the meson fields:
\begin{equation}
\langle X_0 \rangle\ \simeq\  -\ \frac{5 N_m\, |\mbox{Tr}' \lambda|^2\,
  (\mbox{Tr}' \lambda)^\star}{12 (\ln 4-1) h^2 N (N_f-N)}\
  \frac{f^2}{\lambda_{\Sigma} v}\ \ll\ \lambda_\Sigma v\ ,
\label{cw1}
\end{equation}
\begin{equation}
\langle Y_0 \rangle\ \simeq\  -\ \frac{5 N_m\, |\mbox{Tr}' \lambda|^2\,
  (\mbox{Tr}'' \lambda)^\star}{192 \pi^2 N}\
  \frac{f^2}{\lambda_{\Sigma} v}\ \ll\ \lambda_\Sigma v\ ,
\label{cw2}
\end{equation}
where $\mbox{Tr}'' \lambda \equiv {\sum_{i=1}^{N}} \lambda_{X,i}^i$.
The contribution of Eqs.~(\ref{cw1}) and~(\ref{cw2}) to the vacuum energy,
being suppressed both by a loop factor and by
$\langle X_0 \rangle, \langle Y_0 \rangle \ll \lambda_\Sigma v$, is negligible
compared with the ISS energy. Hence, we still have a metastable
vacuum around $\langle \phi \rangle = \langle \tilde{\phi} \rangle = 0$,
with a small tadpole induced for $X$. This plays an important role
in generating $\mu$ and $B \mu$ parameters of the appropriate size
in the MSSM Higgs sector, as discussed in Section~\ref{sec2}.

\end{appendix}

\vfill
\eject



\end{document}